\documentclass[pra,showpacs,twocolumn,showkeys,floatfix]{revtex4-1}
\usepackage{graphics}
\usepackage{amsmath}
\usepackage{amsfonts}
\usepackage{amssymb}
\usepackage{slashed}
\usepackage{mathtools}
\usepackage{color}
\usepackage{bbold}
\usepackage[usenames,dvipsnames]{xcolor}
\usepackage{dsfont}
\usepackage{braket}
\usepackage[latin1]{inputenc}
\usepackage{hyperref}
\usepackage{graphicx}
\usepackage{epsfig}
\usepackage{epstopdf}

\newcommand{\bstfile}{osajnl}

\begin{document}
\newcommand{\bibs}{ajp-pm-paper}

\title{An apparatus for measuring a material's photomechanical response}
\author{Elizabeth Bernhardt, Chad Garrison, Nathan Rasmussen, Joseph Lanska, and Mark Kuzyk}
\email{eabernhardt@wsu.edu, kuz@wsu.edu}
\affiliation{Department of Physics and Astronomy, Washington State University, Pullman, Washington  99164-2814}
\date{\today}

\begin{abstract}
Please find the final and published version of the manuscript using this citation: E. Bernhardt, et. al, ``An apparatus for measuring a material's photomechanical response'',
American Journal of Physics, 86, 14 (2018)

This paper describes a simple and inexpensive apparatus for measuring the light-induced shape change of a material, which can be implemented in a high school or undergraduate laboratory.  The key components are a laser pointer to actuate the material, a force sensor from an inexpensive jeweler's balance to measure the response, an Arduino for data acquisition and a means for mechanically mounting the components.  The apparatus described here was used by high school students and teachers in a summer program to characterize liquid crystal elastomers.  The theory of the photomechanical response is used to interpret the data, from which the material parameters are determined.
\end{abstract}

\maketitle

\section{Introduction}

%

Bringing modern research into undergraduate labs often poses financial and technical challenges.  Furthermore, complex equipment and esoteric theory adds an additional barrier to learning the fundamental principles. Expensive self-contained modern physics experiments can be purchased or built in-house, but the inner workings of the apparatus may be hidden from the student.  This paper describes the design of a simple and inexpensive apparatus for taking research-quality data.  The apparatus can be pre-built for the students and used as a lab module, or its construction could form the basis of a full quarter or semester course, where the students learn about programming Arduinos, building electronic circuits, calibrating sensors, and building an apparatus from the ground up. Salvaging parts from commercial products also emboldens the student to tinker and investigate how things work.

The photomechanical effect is the change in a material's shape (such as a length change) under light exposure.  The apparatus developed here is the clamped configuration, in which a stretched sample is rigidly held in place by its ends, and the light-induced force measured.  While the effect is well known and documented with many materials, a standardized characterization technique has yet to come into common practice.\cite{white2017photomechanical}  Our apparatus is specifically designed to characterize the material's response function.

A senior undergraduate student (second author on this paper) learned the theory of the photomechanical effect, then built the experimental apparatus based on his understanding of how it was to work -- using our research instrument as a guide.  Once built, the apparatus characterized materials, teaching high school students experimental techniques, data collection, error analysis, and fitting data to a model.

\subsection{A brief history of the photomechanical effect}

In the late 1800s, Alexander Graham Bell demonstrated that light modulated with a chopper could induce sound in a thin diaphragm of various materials at the chopper's frequency, thus making the first known observation of the photomechanical effect.\cite{bell1880}  His intention was to make a photophone, where telephone conversations were transmitted on a beam of light.

Since the first photophone, new photomechanical materials have been developed.  The Uchino walker\cite{uchino1990} uses a bi-ceramic leg with one part made of an elecrostrictive ceramic\cite{mason48.01} that bends in response to pulsed light, resulting in ``walking''.  The photomechanical effect of a polymer optical fiber\cite{kuzyk95.03} was put to use by Welker et al to make an intelligent optical/ mechanical transistor\cite{welke94.01,welke94.02} that was miniaturized into a waveguide device.\cite{welke95.01}   Later, Camacho-Lopez et al showed that a curved buoyant elastomer sheet swims on the surface of water to avoid the illumination of a pulsed light source.\cite{camacho-lopez2004}  More such demonstrations can be found in the literature.\cite{white2017photomechanical}

Liquid crystal elastomers (LCEs), composite materials made of multiple components, are often used because of their large photomechanical strain response.  Large light-induced changes in the material's length originates in a two-step process.  In general, the long-range orientational order of the rod-shaped liquid crystal molecules is destroyed by the absorption of a photon, leading to an isotropic orientation of the liquid crystals.  This phase change typically drives a reconfiguration of the soft-rubbery backbone of the elastomer, which is generally chemically connected to the liquid crystals, leading to a large length change.  The details of the theory are well beyond the scope of this paper, but the interested reader is referred to a paper by Harvey and Terentjev, who apply a similar apparatus to study this mechanism.\cite{harve07.01}  The apparatus described here can be used by students to study the mechanisms of the response \cite{corbett2009,camacho-lopez2004, corbett2009, finkelmann2001}, which are critical in applications such as light-controlled artificial muscles \cite{degennes1997, cladis1999, wermter2001} and for understanding new phenomena, such as collective molecular re-orientation as proposed by Bian et al.\cite{bian06.01}
%

\section{Theory of Photomechanical Effect}

\begin{figure}[htbp]
   \centering
   \includegraphics[]{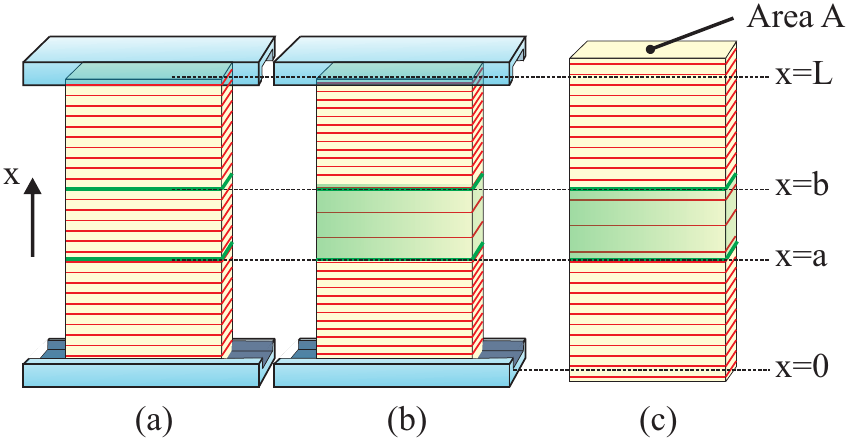} 
   \caption{(a) Fiducial lines (red) are drawn on a sample material (yellow) for reference. (b) The clamps at top and bottom prevent the sample length from changing when light illuminates its mid region (green) and causes expansion. (c) An un-clamped sample expands in response to light.}
   \label{fig:Clamped-Unclamped}
\end{figure}

Appendix \ref{sec:TheoryAppend} elaborates on the theory of a photomechanical response, which one might distill to advanced undergraduate students who are interested in a more rigorous development that allows for a broader variety of measurements.

Figure \ref{fig:Clamped-Unclamped}a shows a sample, clamped at its ends, with no light applied.  The red lines are drawn on the cartoon sample at equal intervals so that deformation can be more easily discerned.  If light causes the illuminated center portion of the sample to expand, as shown in Figure \ref{fig:Clamped-Unclamped}b, the sample in the dark regions will compress to preserve the fixed length, as imposed by the clamps.  Figure \ref{fig:Clamped-Unclamped}c shows an un-clamped sample, which expands by an amount equal to the expansion of the illuminated part.  Note that some materials contract when illuminated.

In a typical experiment, the light is turned on, and the force on the clamps or length change is recorded as a function of time when in the clamped configuration.  While the time-dependence of the resultant photomechanical force is commonly exponential, some materials might show a more complex response.  In most cases, the sample reaches a steady state after a few seconds.  This steady-state response alone yields important information and is measured directly by reading a jeweler's scale attached to one end of the sample or by observing the displacement with a microscopic.  The stress $\sigma = F/A$, defined as the force $F$ per unit cross-section area $A$, is the simplest quantity to measure and is therefore the focus of our studies.

The steady-state stress is a function of the intensity and is well approximated by
\begin{align}
\sigma \approx \kappa_0^{(1)} I + \kappa_0^{(2)} I^2
\end{align}
where the linear term typically dominates, where $\kappa_0^{(1)}$ is the first order photomechanical constant and $\kappa_0^{(2)}$ is the second order photomechanical constant. These constants characterize the photomechanical response. A large constant, on the order of 100 s/m, indicates a large stress response. The constants are chosen to be positive for a length increase and negative for a length decrease. Determining the sign of the constant is discussed in the following section.

This is the minimal theory needed to do simple experiments that characterize  $\kappa_0^{(1)}$ and $\kappa_0^{(2)}$, and a jewelers scale can be used directly when the time dependence is not required. For this simple study, a binder clamp available at an office supply store can be epoxied directly to the scale that holds one side of the sample, where the other side is similarly clamped to a fixed support.  Then, the display is read directly by the experimenter as she illuminates the sample with a laser pointer held by hand, and the steady-state reading is noted for analysis.  This was the first experiment that we tried at the start of our studies and might be a good first step for those wanting to gain experience before building a more sophisticated apparatus.    Section \ref{sec:DataAnal} describes our analysis of the time dependence of the stress and Appendix \ref{sec:TheoryAppend} describes how more complex response functions can be determined. 

\section{Apparatus design}

This section describes the hardware and software designs for the apparatus. Hardware used in our apparatus are listed in Table~\ref{table:hardware_table} with approximate price ranges.  Suggestions for less expensive replacements are provided in the text.
\subsection{Hardware}
\begin{table}
\begin{tabular}{l c r}
\hline
Item & Distributor & Price\,(USD)\\
Arduino & Amazon & 11-30\\
INA125 operational amplifier & Mouser & 6-7\\
Basic electronic components* & Amazon & 10-90* \\
Jeweler's scale & Amazon & 10-15 \\
Calibration weight* & Amazon & 5-10*\\
Optomechanical components* & Thorlabs& 130-2000*\\
Optical breadboard* (1\,ft by 1\,ft)& Thorlabs &151-250*\\
Laser pointer & Amazon& 5-50\\
Neutral density filters* &Thorlabs& 460-650*\\
Power meter & Thorlabs &100-500\\
Scrap aluminum* & Scrap yard & 10-50*\\
\hline
Total (exact replica) & &898-3652\\
*Total (less expensive options) & &132-602\\
\hline
\end{tabular}
\label{table:hardware_table}
\caption{Experimental components for the photorheometer outreach setup. An Arduino may be purchased with a starter electronics kit. Optical components were priced individually for the lower bound and priced as kits for the upper bound. A larger than necessary optical breadboard was selected to maximize flexibility in the setup. Neutral density filters may be replaced with inexpensive cellophane. Most items from Thorlabs may be sourced from Ebay or other used goods retail services, including power meters. * indicates items are optional and less expensive options are discussed in the text.}
\end{table}

\begin{figure}[htbp]
   \centering
   \includegraphics[]{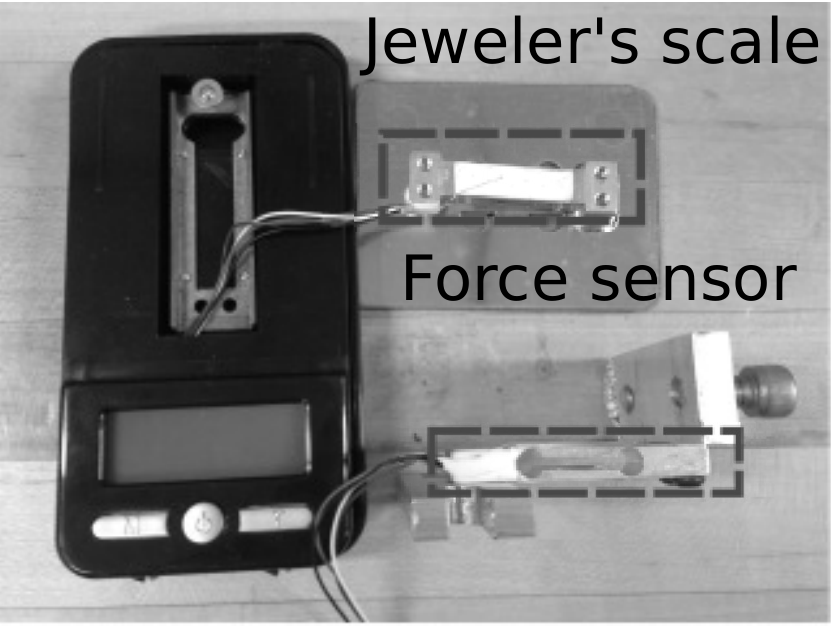}
   \caption{The strain gauge (within dashed lines) is salvaged from a jeweler's scale.  To the right is a strain gauge still attached to the bottom of the pan. On the bottom right is a strain gauge which has been attached to mounting hardware made from scrap aluminum.  Note the wiring in the strain gauge is delicate and is easily damaged.}
   \label{fig:force_sensor}
\end{figure}

The primary components of this apparatus are an Arduino UNO microcontroller and a force sensor.  Arduino products are inexpensive open-source microcontrollers with several analog and digital input and output pins, can execute code the user uploads to the board, and provide a 5\,V power supply.  The Arduino UNO is widely used, as it is perfect for beginners to electronics and coding.\cite{ArduinoWeb} In fact, many apparatus designs featured in the American Journal of Physics are based on Arduino technology. The Arduino UNO may be sourced on its own from various retailers or can be purchased in a beginner's Arduino kit, which includes basic electronic prototyping equipment.

The strain gauge is harvested from a low-cost jeweler's scale; see Figure~\ref{fig:force_sensor}. Any jeweler's scale may be substituted; our scale is an American Weigh Scale Digital Pocket Gram Scale, measuring a maximum of 100\,g in 0.01\,g increments. Note the precision of the jeweler's scale is irrelevant, as the strain gauge is completely removed from the jeweler's scale electronics. As such, the least expensive jeweler's scale one can find should be substituted. In this paper, strain gauge and force sensor are used interchangeably.

A typical strain gauge has a red wire to provide the voltage supply, a black wire to ground the sensor, and two signal wires that are typically referred to as sensing voltages $S^+$ and $S^-$. In Figure~\ref{fig:force_sensor}, the white and green wires are the sensing voltages for the bottom force sensor, while the white and blue wires are the sensing voltages for the top sensor. The force sensor outputs a voltage that encodes the force, and the Arduino reads this voltage.

Our setup is made from machined aluminum parts, but wood or household items could be used as substitutes.  Sample-holding clamps are milled into three pieces: a rectangular base with threading for screws (4\,cm width, 1.85\,cm height, 6.5\,cm thickness), an aluminum bar with two holes (4\,cm width, 1.85\,cm height, 1.75\,mm thickness), and screws to attach the bar to the base and the base to other components. The distance between the screws determines the maximum width of the sample; our clamps have 2.5\,cm clearance between screws. As such, wider clamps may be machined to hold wider samples. Each clamp has an aluminum bar, and it holds the sample using two screws, which press on the bar. Instead of aluminum, the clamps may be replaced with plastic office binder clips if access to machining equipment is limited.  The clamping method is not critical as long as it is rigid and does not allow for the sample to slip.

\begin{figure}[htbp]
   \centering
   \includegraphics[]{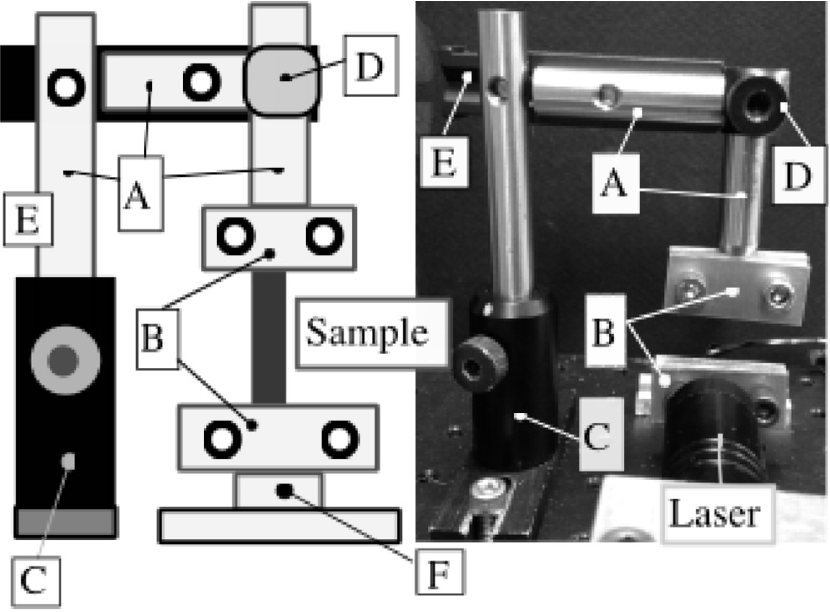}
   \caption{Left: diagram of apparatus as viewed from the front with sample in between clamps; Right: photo of apparatus without a sample between clamps. (A) optical posts (B) custom machined sample clamps (C) post holder with thumb screw on post mount attached to optical table (D) fixed 90$^\circ$ angle post mount with thumb screw attached to horizontal optical post and vertical optical post (E) mounting joist to create 90$^\circ$ angle between vertical optical posts and horizontal optical post (F) force sensor with one edge attached to clamp and opposite edge attached to post mount}
   \label{fig:setup_frontview}
\end{figure}

The bottom clamp is fixed to the force sensor.  The two threaded holes in the force sensor may be drilled out to make clearance holes for mounting, which prevents cross-threading.  Extreme care must be taken to protect the wiring, as even minor damage to the wires could cause the entire sensor to malfunction. The other end of the force sensor is attached to the optical table. The top clamp is attached to a vertical translation system. Figure~\ref{fig:setup_frontview} shows a manual vertical translation system constructed using ThorLabs optical post assemblies. However, optomechanical products may be cost prohibitive. Using a handheld drill, one may drill a small hole in the plastic binder clip, then use washers and screws to attach the clip to scrap wood, aluminum, or other more readily sourced materials. A note of caution: attaching clamps to posts with superglue or other glues is not recommended, as the contact point may be more flexible than using metal screws, which may lead to the apparatus measuring the response of the glue to the photomechanical force provided by the material, instead of measuring only the material's response. Screws may be harvested from the jeweler's scale, and washers may be created from the scale's outer plastic casing.

The apparatus is built on an optical breadboard for portability and because the breadboard was readily accessible in our lab. The breadboard may be replaced with any suitable flat-surface item, such as a plywood sheet, wood from a pallet, or a household baking sheet. Attaching the apparatus to any breadboard replacement item may require handheld power tools to drill pilot holes so the clamps, laser holder, etc.\ may be attached using screws.

\begin{figure}[htbp]
   \centering
   \includegraphics[]{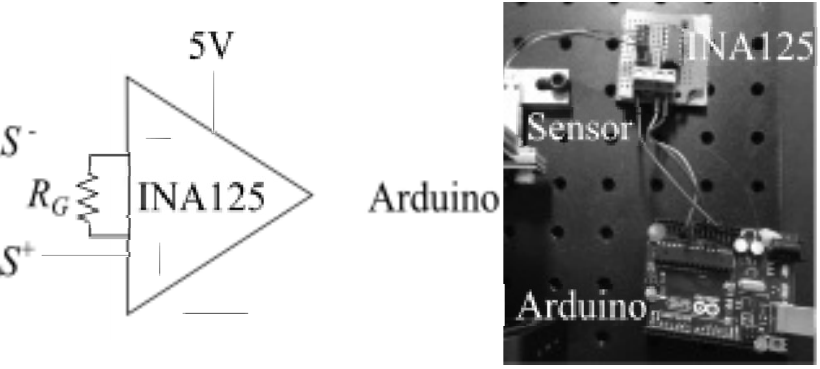}
   \caption{Left, amplification circuit; right, circuit connected to sensor and arduino. The sensing voltages, $S^+$ and $S^-$, are amplified using a variable gain operational amplifier (op-amp), INA125. The gain is set with the external resistor $R_G$. This resistor is not connected to the op-amp inputs but rather to pins~8 and~9.}
   \label{fig:circuit}
\end{figure}

A circuit, shown in Figure~\ref{fig:circuit} and connected to the outputs from the force sensor, amplifies the signal and provides an electronic buffer between the scale and the Arduino. The circuit is necessary as the voltage reading from the strain gauge is on the order of microvolts; changes in this reading are difficult for the Arduino UNO to detect without amplification. The selected integrated circuit is a low supply-power operational amplifier, INA125. This operational amplifier was chosen because its supply voltage requirements match the voltage supplied by the Arduino UNO. While the amplification circuit may be run off a separate, non-Arduino power supply, the apparatus was designed to be self-contained and powered by the Arduino. The circuit amplifies the difference between the sensing voltage signals from the force sensor according to
\begin{eqnarray}
V_o &=& \left( S^+ - S^- \right)G,\\
G &=& 4 + \frac{60 \mathrm{k}\Omega}{R_G},
\end{eqnarray}
\noindent where $R_G$ is a resistor externally connected to the INA125 chip. The maximum amplification is 10,000; thus the external resistor has a lower limit of $6\Omega$. In our experimental setup, $R_G$ is a $100\Omega$ potentiometer, allowing for the gain to be adjusted between $G=604$ and $G=10,000$. Potentiometers with such a low resistance are expensive and difficult to find, so a fixed gain circuit may be created instead. Higher gains yield higher sensitivity of the force sensor reading, but the system noise is also amplified. The circuit output connects to one of the analog pins of the Arduino based on the user defined executable code, which is described later.

A laser pointer photo-actuates the sample, and its wavelength must be selected based on the sample's absorption spectrum. A green laser pointer with $\lambda$ = 514\,nm was found to work with samples using azobenzene dyes. Other wavelengths of laser pointers may be used as well; the high school students tested their materials with a 445\,nm laser. Instead of relying on battery power, the laser pointer was powered with a DC power supply. Caution must be taken when running the laser pointer for extended periods of time, regardless of how the laser pointer is powered. Most laser pointers do not have appropriate heat sinks for the heat energy generated by running the laser pointer continuously for more than 5-30 seconds. Our students circumvented the heatsink problem by blowing canned air onto the laser pointer. However, the 445\,nm laser pointer still failed after one experiment, rendering it unable to lase. Subsequently, a transistor was used to turn the power to the 514\,nm laser on and off using an Arduino.

\begin{figure}[htbp]
   \centering
   \includegraphics[]{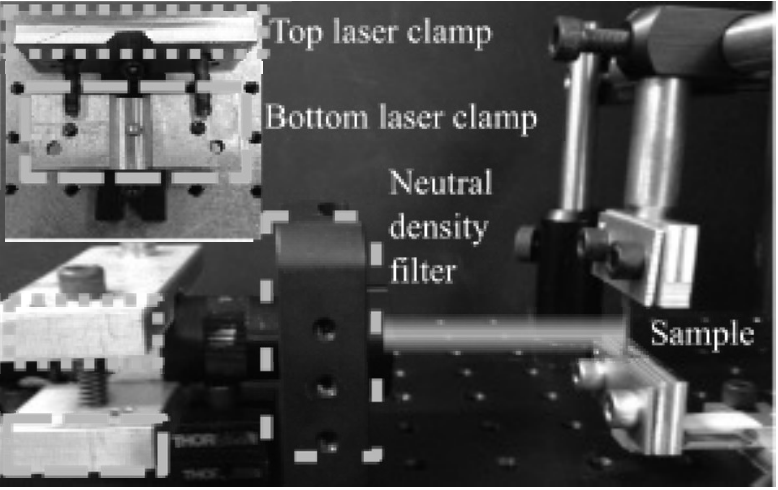}
   \caption{Side view of the setup with neutral density and its holder placed in the beam's path, shown by the solid green line. The inset shows grooves machined out of repurposed aluminum parts to form clamps for holding the laser. Dimensions are not given as the grooves are dependent on the laser's diameter.}
   \label{fig:setup_sideview}
\end{figure}

Figure \ref{fig:setup_sideview} shows the laser holder, which is made of two custom machined aluminum clamps. The bottom clamp piece attaches to the optical table with two optomechanical components underneath to increase the height of the laser; note that in the inset to Figure~\ref{fig:setup_sideview}, a screw is shown to be threaded into the bottom laser clamp. The top clamp attaches to the bottom clamp using two long screws, which are adjusted until the laser is secure. The laser may be held by the clamp or by hand. Other options include o-ring clamps attached to scrap wood or aluminum. Before our laser holder was machined, our laser pointer was held in place using duct tape and scrap aluminum, which was then duct taped to the table.

Various optical density filters may be set in the beam's path to adjust the intensity of the beam.  Inexpensive colored plastic sheets or polarizing sheets can be used if on a budget. Additionally, a transparent jar filled with water may be used as a substitute optical density filter; appropriately colored food coloring is added to the water in increasing amounts so that the water absorbs some of the light passing through the jar. Online photography resources on homemade neutral density filters are another good resource for do-it-yourself intensity adjustment optics; one such resource suggests welding glass.\cite{photographyweb} If desired, beam shaping can be achieved using various optics, such as cylindrical lenses, or razor blades may be used to create a rectangular beam profile.

\subsection{Software}\label{subs:software}
Our custom Arduino code is loaded on to the force sensor in two steps, though an undergraduate proficient in coding may create one code capable of performing both functions. The first code zeroes the scale, and the second code streams the data. Codes and their usage will be covered in Section~\ref{subs:datac}.

The Arduino board is designed to stream data to a computer through a USB link, but it is a bit more difficult to write this stream to a file using only Arduino sketches. Our proposed workaround uses Python and a virtual serial port. Python is an open-source programming language with many forums, packages, and software resources available to the user.\cite{PythonWeb} One such package is PySerial, which provides a back end for interfacing serial ports.\cite{PySerialWeb} Using PySerial, the data are streamed and written to a file.

Stored data is first transformed from raw force measurements to stress with custom Python code, then analyzed with Origin Pro data analysis software. Any software capable of fitting data would be appropriate.
\section{Experiment}
The laser is passed through spatial beam shaping optics and a neutral density filter to control intensity. The area of the beam was measured using a BeamStar beam profiler, but beam profilers are expensive so a razor blade and power meter can be used to determine the beam size; see Ref.~\cite{galvez2006gaussian}.  A ThorLabs DET110 high speed silicon detector is used to measure the laser power. From these measurements, the intensity is calculated using $I=P/A_{beam}$.  Alternatively, an inexpensive photocell and voltmeter can be used, depending on the budget.

A sample is placed in the top clamp and the screws tightened, then lowered into the bottom clamp and its screws tightened. If the sample is buckled, the top clamp is slightly raised to straighten the sample.  The length between the clamp, width, and thickness of the sample are then measured and recorded.
\subsection{Suggested samples}
Poly(methyl methacrylate) (PMMA) doped with azobenzene dyes is a good material system for studying the photomechanical effect. While PMMA is commercially available as Plexiglas, Acrylite, and Lucite, creating usable samples from these commercial products is challenging. Polystyrene, commonly available as uniformly pre-strained sheets called Shrinky-Dinks, reacts to light, but it is usually an irreversible effect due to relaxation of production-induced strain.  Colored cellophane can also be used.

Dye-doped liquid crystal elastomers were created by the high school students using a two-stage synthesis developed by Yakacki et al.\cite{saed2016} The first stage is a Michael-thiol addition, and the second stage aligns the liquid crystal molecules permanently using UV light activated physical cross links. The synthesis is easy enough for the high school students to create their own samples, yet flexible enough for them to select a particular dopant concentration. Disperse red 1 acrylate dye (DR1) was used as a dopant for its known photomechanical effect. While this synthesis method was performed by our high school students, they had access to the necessary materials and synthesis apparatuses, as well as training and proper safety equipment. The majority of physics laboratories may not have access to chemistry labs, so this may provide an opportunity for interdisciplinary collaborations with one's chemistry colleagues.  A few more readily accessible material options are described below.

First, we tested commercially available PMMA fibers doped with fluorescent dyes from Fiber Optic Products Inc.\cite{FiberOpticsInc} Fibers from this company may be purchased in a variety of colors and dimensions; they are made of a polystyrene core with an acrylic cladding, making them ideal candidates for this experiment. We tested yellow fibers with a square 1.5\,mm by 1.5\,mm cross section (product code YS1500), as this geometry is easier to model with more advanced theory. Green square fiber may be useful as well (product code GS1500). Results for the yellow fiber are presented in the results section.

\begin{figure}[htbp] 
   \centering
   \includegraphics[]{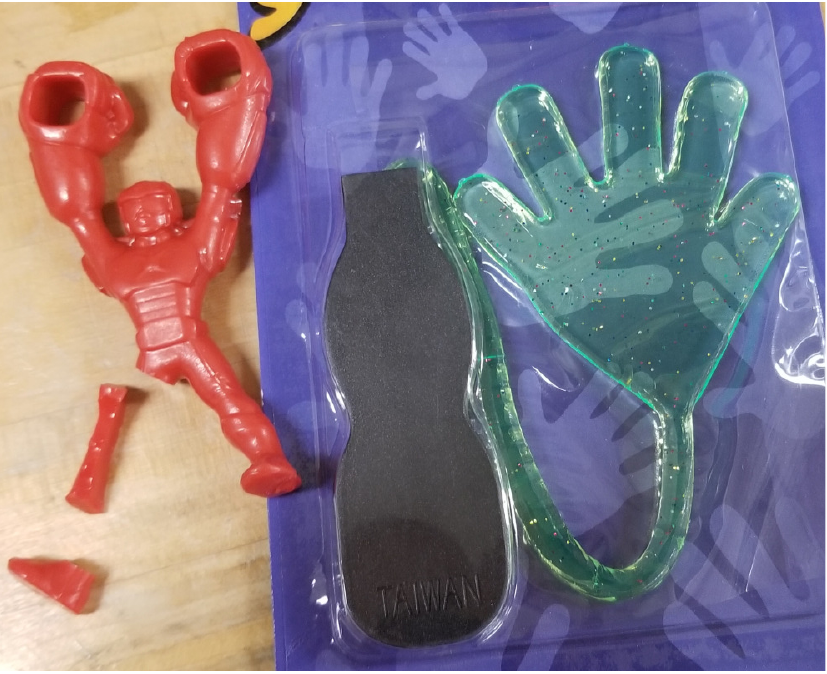}
   \caption{Toys from a local dollar store which were tested for a photomechanical response. Left: flying superhero toy with leg cut to create a sample to test. Right: Blue sticky hand throwing toy. Note a pink sticky hand throwing toy was tested in this paper.}
   \label{fig:toy_photo}
\end{figure}

Additionally, we tested two toys sourced from a local dollar store: a sticky hand throwing toy and a flying superhero toy (see Figure~\ref{fig:toy_photo}), as well as a painted rubber band. A suitable search criteria for identifying good materials includes lightly crosslinked dye-doped elastomeric materials, which are identifiable by stretchiness, reversible deformation, and coloring that is absorbing to light. A rubber band is a commonly available item that is lightly crosslinked. These lightly crosslinked materials are stretchy, like a sticky hand throwing toy, but return to their original shape after deformation, like a flying stretchy superhero toy. A material that is stretchy but does not return to its original shape, like a gooey stress-relief non-Newtonian fluid toy, would not be ideal for this experiment. Finally, the color of the material should, ideally, absorb the wavelength of the laser light one intends to use to test the material. A rubber band was painted with black spray paint to maximize absorption of the light. In general, red materials absorb blue light strongly, so the red flying superhero toy and a pink sticky hand were tested with the 488\,nm laser. One should consult elementary texts on color absorption and reflection to gain an intuition on suitably colored materials for the laser light one intends to use.  However, the general ideas presented here on material and laser selection are enough to make the experiments work. Results for the toys and rubber band are presented in the results section.
\subsection{Data collection}\label{subs:datac}
Data collection occurs in two stages. First, the setup is calibrated after sample placement using an Arduino calibration code. Next, the calibration constants are input into a second Arduino code. Finally, the stream from the Arduino is captured by a Python script and recorded.

Prior to data collection, the Arduino code is run for calibration as discussed in Section \ref{subs:software}, which takes less than a minute.  The zero reading corresponds to the number streamed from the analog Arduino pin when the sample is secured to the setup, straightened, and, optionally, pre-strained.  A 50\,g calibration weight is set on the bottom clamp, and the reading from the Arduino is taken. If the reading from the 50\,g weight is lower than the zero reading, the force sensor has been installed upside down and must be reinstalled; alternatively, the calibration constants may be set to -50 for the low value and 0 for the high value. A nickel or other form of coin currency may be used in place of a calibration weight, as the mass and its tolerance values are governed by federal laws. These two measurements are input into the second Arduino code, then this second code is loaded to the Arduino. Now, the Arduino is calibrated and ready to collect data. The Arduino must remain connected to the computer via the USB link during data collection.

When ready to take data, the custom Python script is executed during the entire data collection process and only stopped when data collection is completed.  For optimal data streaming, the hard drive is kept running during data collection.

\begin{figure}
   \centering
   \includegraphics[]{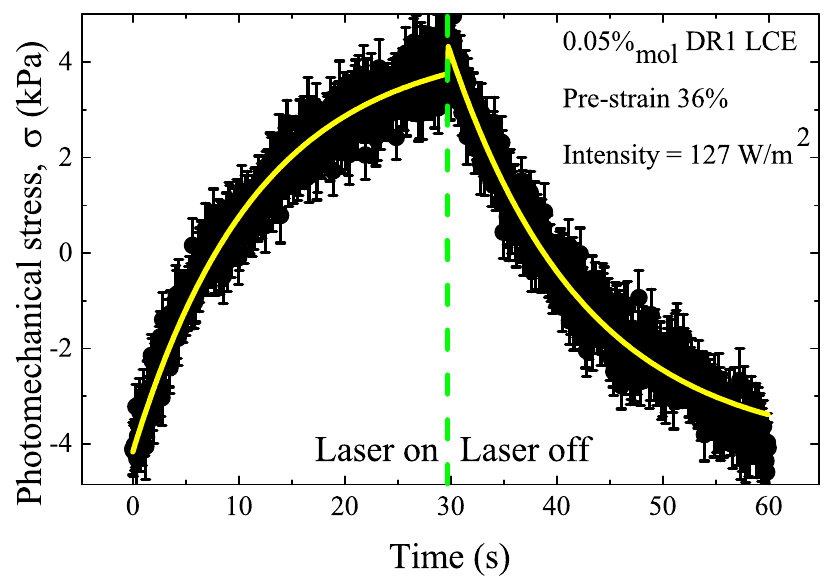}
   \caption{Representative LCE photomechanical stress response data from one on-off cycle (black data points) and exponential fit (light curve). The laser is turned on at $t=0$ and turned off at $t=30$\,s. Each point is an average over 25 voltage readings from the force sensor. The sample first decreases in length, as indicated by an increase in the measured stress.}
   \label{fig:example_data}
\end{figure}

The laser illuminates the sample for a measured amount of time, $t_{\mathrm{off}}$, which is based on the relaxation time constants of the material.  In general, stiffer materials tend to have shorter time constants and less stiff materials have longer time constants.  After the elapsed time, the laser is blocked for a fixed amount of time, completing one on-off cycle. The laser can be blocked by hand, or with a servo motor and a second Arduino to create a shutter.  The process is repeated to test repeatability and allow for data averaging.  One such on-off cycle is displayed in Figure \ref{fig:example_data}. In this run, students stretched the material (labeled pre-strain in the figure), then measured the photomechanical effect at various intensities.  One intensity is shown. For fibers or materials with a high Young's modulus, 15 seconds on and 15 seconds off suitably captured the photomechanical effect. Ten on-off cycles were collected per intensity, so the experiment took 5 minutes per intensity. For elastomers or materials with a lower Young's modulus, 30 seconds on and 30 seconds off, in general, captured the photomechanical effect, so the same experiment took 10 minutes per intensity. Five to ten different intensities are measured, based on how much time the student has to collect data.

To determine $\kappa^{(i)}_0$, the stress on the sample is measured as a function of intensity using neutral density filters in beam's path to change the intensity starting with a high optical density filter, thus working from a low intensity to a high intensity by changing the filters and measuring and recording the intensity. The beam is blocked when changing the filter.  Samples may become photodamaged if the laser intensity is too high, so samples should be inspected either during the blocked part of the on-off cycle or while changing optical density filters. Photodamage is signified by a change in color or holes in the sample.

\subsection{Data analysis}\label{sec:DataAnal}
The end goal of data analysis is to find the photomechanical constants, $\kappa_0^{(i)}$. The first step transforms raw data into stress as a function of time data ($\sigma(t)$). Next, this function is fit using appropriate exponentials, which produces fit constants. Finally, the fit constants are collected to create $\bar{\sigma}(I)$, which is the equilibrium stress as a function of intensity, and this final function is fit to determine $\kappa_0^{(i)}$. In this section, these steps are detailed.

Our data collection Python script creates a file with elapsed time $t$ in one column and the measured ``force'' $f$ in grams in the second column. The stress on the sample is given by
\begin{eqnarray}
\sigma \left( t\right) & = & \frac{f\left( t \right) G}{t_0 w_0},
\end{eqnarray}
where $w_0$ is the width of the sample, $t_0$ is the thickness, and $G$ is the gravitational constant, which is necessary to convert the collected ``force'' reading into a physical force.  A 25-point average is performed to improve noise and the standard deviation is used to determine the uncertainty.

At a given light level, a sample may not respond at all, may increase in length -- pushing on the force sensor thus decreasing the output voltage, or may decrease in length -- pulling on the force sensor and increasing the voltage reading.  Depending upon how the sample responds, the signal will either increase or decrease back to the baseline when the light is turned off.

For a sample that decreases in length under illumination, the data is fit to the exponentials
\begin{equation}
\sigma \left( t, I \right) =
\left\{
\begin{array}{ll}
      \sigma^{\mathrm{on}}_{eq}\left(I\right) \exp{\left(-\frac{t}{\tau_{\mathrm{on}}}\right)} + \sigma_0 & t\le t_{\mathrm{off}} \\
      \sigma^{\mathrm{off}}_{eq}\left(I\right)\left(1-\exp{\left(-\frac{t-t_{\mathrm{off}}}{\tau_{\mathrm{off}}}\right)} \right) +\sigma_0 & t > t_{\mathrm{off}} ,\\
\end{array}
\right. \end{equation}
where $\sigma^{\mathrm{i}}_{eq}\left(I\right)$ is the stress at $t=\infty$ for an illuminated sample ($\mathrm{i}=\mathrm{on}$) and ($\mathrm{i}=\mathrm{off}$) for a dark sample. $\tau_{\mathrm{i}}$ is the time constant in seconds, which may be a function of intensity, $t_{\mathrm{off}}$ is the time in seconds when the laser is blocked, $\sigma_\mathrm{0}$ is offset due to pre-strain, improper calibration of the force sensor, or creep of the sample over time.  Should the sample increase in length, simply switch the equations so that the exponential decay is for $t<t_{\mathrm{off}}$. For each intensity, $n$ fit constants will be collected if $n$ on-off cycles are performed, represented with subscript $i$ (typically, $n=10$). Typically, the fit constants are averaged using a weighted average, detailed in Appendix~\ref{sec:weightedavg}, resulting in the function $\bar{\sigma} \left(I\right)$. One may eliminate the weighted average and fit $\sigma_{i} \left(I\right)$ instead, where $i$ includes all calculated fit constants.

Our interest lies in finding $\bar{\sigma} \left(I\right)$, plotting this function, and fitting it to find $\kappa_{0}^{(i)}$. The most general model is a simple polynomial
\begin{eqnarray}
\bar{\sigma} \left(I\right) & = & \sum_{i=1}^{\infty}\kappa_{0}^{(i)} I^i,
\label{eqn:fit_sum}
\end{eqnarray}
where the summation limit is typically truncated to 1, 2, or -- in rare cases -- 3. In general, the higher the Young's modulus of a material, the more linear the photomechanical response. As such, stiff glassy polymer fibers, such as PMMA fibers, need one photomechanical constant for a satisfactory fit of the data, while gooey liquid crystal elastomers are better described by two or three photomechanical constants.
%

\section{Results}

\begin{table}
\begin{tabular}{l r}
\hline
Sample & $\kappa_0^{(1)}$ (s/m) \\
\hline
DR1-doped LCE, 36\% strain, advanced setup &$-11\pm1.8$\\
DR1-doped LCE, 36\% strain, simple setup &$  -3.2\pm$1.3\\
Fiber Optic Products Inc &$12.4\pm0.4$\\
Black painted rubber band &-0.14$\pm$0.04 \\
Sticky throwing hand toy & N/A \\
Flying superhero toy &$-0.44\pm0.02 $\\
\hline
\end{tabular}
\label{table:kp_all}
\caption{A summary of samples tested for the photomechanical response. The DR1-doped liquid crystal elastomer was tested in both the advanced setup and the simple setup, where the wavelength of the stimulating light was different, among other apparatus changes. More commonly available samples were subsequently tested with the advanced apparatus, and the results are summarized.}
\end{table}

\begin{figure} 
   \centering
   \includegraphics[]{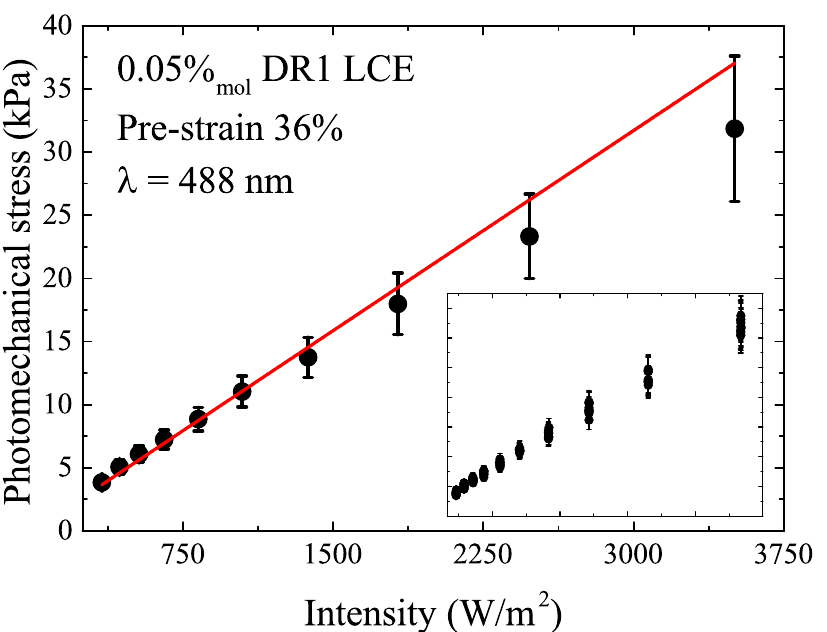}
   \caption{Representative graph to find $\kappa_0^{(1)}$ for DR1-doped liquid crystal elastomers using the advanced apparatus.  This systematic deviation from linearity at higher intensities may be due to the nonlinear contribution, which we did not study here. The inset shows all the calculated fit constants for each on-off cycle (ten on-off cycles were collected per intensity), which were averaged using a weighted average to produce the ten data points in the main graph.}
   \label{fig:kp_graph}
\end{figure}

First, we compare the results from the outreach setup used by high school students with the more advanced apparatus used in our lab. Figure \ref{fig:kp_graph} shows results from the advanced setup, which includes an adjustable offset, adjustable gain, secondary amplification circuit, full automation, a stable 1\,W Argon-Krypton laser at $\lambda$ = 488\,nm that illuminates the whole sample, and a built-in Young's modulus testing cycle. In contrast, the outreach setup uses a $\lambda$ = 514\,nm green laser pointer with a significantly lower maximum power output, which focuses to a small portion of the sample to increase the intensity to the same order of magnitude as the research-grade apparatus.

The sample used in both setups is a DR1 (disperse red 1) acrylate dye-doped liquid crystal elastomer (LCE) synthesized by the high school student team. The sample has a concentration of $0.05\%_{mol}$ and has been pre-strained to 36\%. Note the data trend for the advanced setup is linear and yields $\kappa_0^{(1)} = -11\pm1.8$\,s/m. While a second or third order polynomial may seem to better fit the data, error analysis demonstrates a linear fit best fits this data. The negative sign indicates the sample decreases in length when exposed to light.

\begin{figure}
   \centering
   \includegraphics[]{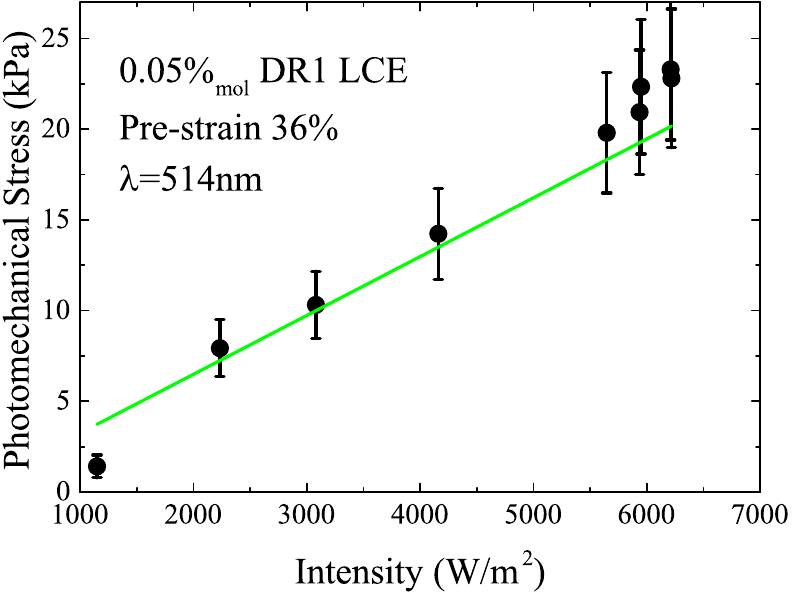}
   \caption{Data collected from the outreach setup to find $\kappa_p$ with weighted average $\bar\sigma$. Similar to Figure~\ref{fig:kp_graph}, 10 on-off cycles were collected at each intensity, and the fit constants were weighted using a weighted average.}
   \label{fig:setup_graph}
\end{figure}

Data from the outreach setup are presented in Figure \ref{fig:setup_graph}. The intensity is calculated by monitoring the power of the laser pointer, then dividing by the area of the beam. Using Equation \ref{eqn:fit_sum} gives $\kappa_0^{(1)}= -3.2\pm$1.3\,s/m when limiting the upper limit to $n$=1. The outreach setup is expected to produce a lower photomechanical constant because the excitation wavelength is further from the peak absorption wavelength of DR1 acrylate. While the photomechanical constants determined by the inexpensive setup and the research instrument do not agree within experimental uncertainties, they are in reasonable agreement when taking into account the different measurement wavelength and the illumination geometry of the the sample.

\begin{figure} 
   \centering
   \includegraphics[]{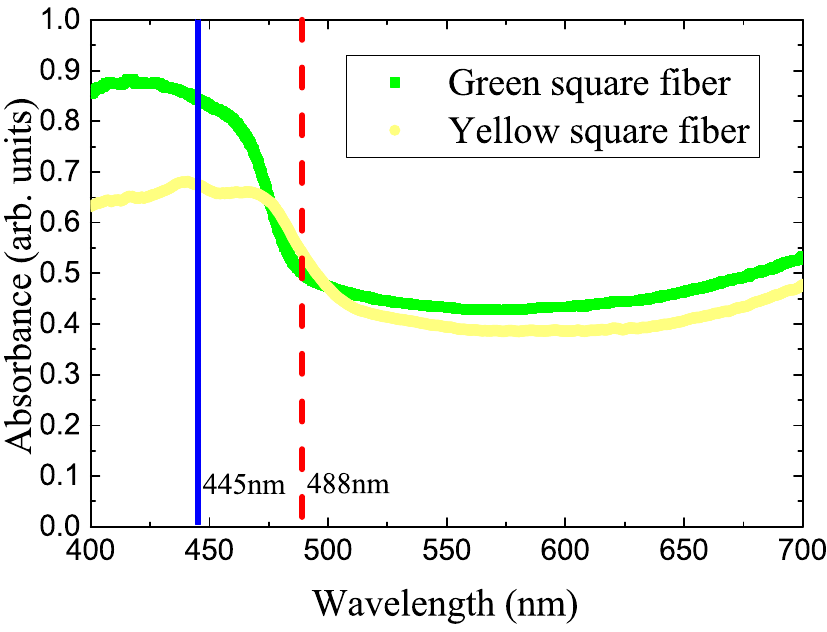}
   \caption{Absorption spectra for two Fiber Optics Inc dye-doped fiber samples with a square cross section. Two pump wavelengths are marked; 488\,nm was used to characterize the yellow fiber while 445\,nm is a readily available laser pointer wavelength.}
   \label{fig:spectra}
\end{figure}

Next, we evaluate readily available commercial products using the photorheometer. Fiber Optic Products Inc.\ fibers were purchased and tested. The rectangular cross section fibers have a high absorbance at the marked potential pump wavelengths, which can be seen in Figure~\ref{fig:spectra}. To demonstrate the photomechanical effect in these fibers, the samples were placed in the advanced setup with the 488\,nm laser to maximize the laser intensity. A laser pointer with a pump wavelength of 445\,nm is suggested for the outreach setup. The outreach setup was first used to test these fibers at 445\,nm, but, as noted previously, the laser quickly overheated from continuous use.
\begin{figure}
   \centering
   \includegraphics[]{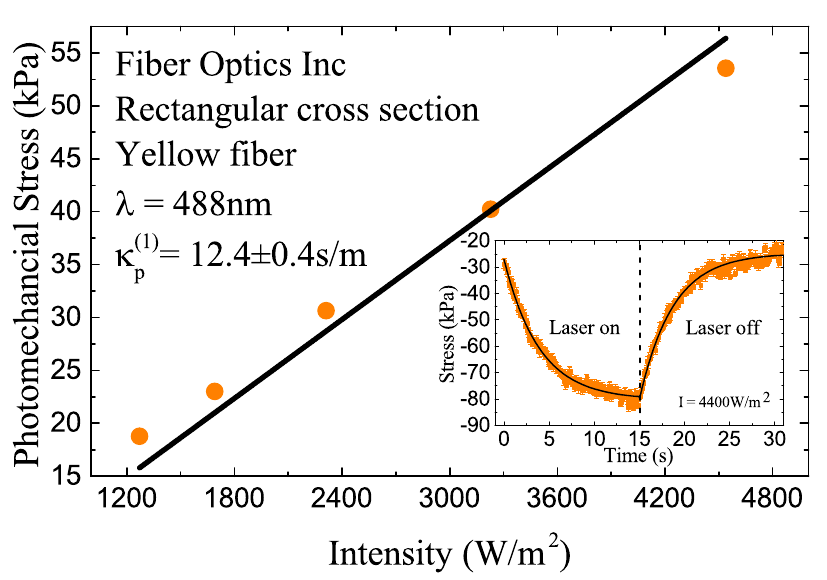}
   \caption{Data collected from the advanced setup to find $\kappa_0^{(1)}$ with a pump wavelength of 488\,nm for a yellow Fiber Optic Products Inc.\ rectangular cross section fiber. Inset shows one on-off cycle for the sample at the highest tested intensity.}
   \label{fig:foi_results}
\end{figure}
Figure~\ref{fig:foi_results} shows the results for the five tested intensities, which have been fit to a line to produce $\kappa_0^{(1)} = 12.4\pm0.4$\,s/m. Each data point in the plot is a weighted average of ten on-off cycles; the inset shows one such on-off cycle. The fiber first increases in length when exposed to light, which can be seen by the decrease in stress in the inset of Figure~\ref{fig:foi_results}. When the laser light is blocked, the fiber relaxes back to its initial length. Notice that the time dependence of the inset is different from the example fit in Figure~\ref{fig:example_data}.

Similarly, we painted a rubber band with black spray paint and placed the sample in the setup. The sample was not strained. Using 488\,nm stimulating light, the rubber band showed a photomechanical response similar in shape to the Yakacki synthesis DR1-doped liquid crystal elastomers in that the sample decreased in length during light exposure, and relaxed back to its original length when the light source was blocked. The data fit to a line and resulted in $\kappa_0^{(1)}=-0.14\pm0.04$\,s/m. The magnitude of the response was small and noisy compared to other samples.

Finally, we tested two toys sourced from a local dollar store. Each toy was cut to a small rectangle using a razor blade, then placed into the setup. The sticky throwing hand showed no photomechanical response, most likely due to the low crosslink density, evidenced by the fact that the sticky throwing hand stretches nearly 200\% its initial length. When a material exhibits a photomechanical response, the underlying mechanism involves dye molecules converting photon energy into thermal or mechanical energy.\cite{white2017photomechanical} Local distortions in a polymer matrix from the dye molecules are translated to a macroscopic deformation as the number of dye molecules participating in the photomechanical effect increases. For a highly crosslinked sample, the microscopic changes are more readily communicated throughout the polymer network. For a lightly crosslinked sample, such as the sticky hand toy, microscopic changes only affect nearest neighbor polymer chains.

However, the superhero flying toy did exhibit a photomechanical effect, with $\kappa_0^{(1)} = -0.44\pm0.02$\,s/m. This sample is heavily crosslinked, as demonstrated by the limited strain observed when the toy is stretched. Moreover, this toy is opaque, whereas the sticky hand was translucent, so all of the light illuminating the sample is absorbed. Therefore, more dye molecules may contribute to the photomechanical effect and the local distortions from the dye molecules is easily translated to the bulk sample via the crosslinked network.

\section{Conclusion}

This paper outlines an economical plan for making an apparatus that measures the photomechanical effect.  The components are harvested from inexpensive commercial products, making the construction of the experiment an educational experience.  The measurement of the photomechanical response in the time domain provides a full characterization of the material response function, which can be used to test various models of the underlying mechanisms.  As such, the availability of such an apparatus in a high school or undergraduate lab enables the student to carry out experiments that are of current interest to the research community.  The simpler apparatus is shown to provide data of sufficiently good quality to compare materials for their response time and strength.  

\section{Acknowledgements}

The authors would like to thank the 2017 STEAM team (Amaya Aranda, An\"{a}is Bautista, Shirley Chen, Marco Miller, Nicholas Abruzzo, Julio San Augstin, and Lizabeth Young) for creating the samples used in this paper. We also thank our generous reviewers for their highly valued input. Support for this project comes from NSF EFRI REM ODESSEI 1332271-4.

\bibliography{\bibs}
\bibliographystyle{\bstfile}

\appendix

\section{Theoretical Details}\label{sec:TheoryAppend}

There are two experimental configurations for measuring the photomechanical response of a material.  Figure \ref{fig:Clamped-Unclamped}a shows the clamped configuration, where the ends of the material are fixed to keep its length constant.  Lines are drawn on the material for reference.  Figure \ref{fig:Clamped-Unclamped}b shows the material when it is exposed to light at its middle section, as shown by the green lines at x=a and x=b.  In the exposed region, the material expands, causing the material in the dark region to compress.  In the un-clamped configuration, as shown in Figure \ref{fig:Clamped-Unclamped}c, the midsection expands but the dark regions remain unaffected.

Ignoring the effects of gravity, a force must be applied to the sample's ends to prevent its total length from changing.  This restraining force is equal in magnitude and opposite in direction to the force that the illuminated region applies to the dark one if the dark region is to remain static.

In the un-clamped configuration, the illuminated region expands unimpeded so that the displacement of the sample ends equals the displacement of the ends of the illuminated region.  As such, the clamped experiment measures the photomechanical force while the un-clamped one determines the displacement.  Past experiments have measured the un-clamped photomechanical response (see, for example, the work of Dawson\cite{dawso11.01,dawso11.02} and Bian\cite{bian06.01}).  The present experiment focuses on the clamped configuration.

Material properties are best described by stresses and strains in lieu of forces and displacements.  The stress and strain are defined by the force per area and the change in length per unit length, respectively, or

\begin{align}\label{eq:stress}
\sigma = F/A
\end{align}
and
\begin{align}\label{eq:strain}
u = \delta L/L ,
\end{align}

\noindent where the stress $\sigma$ is the force $F$ applied to the cross sectional area $A$, and $\delta L / L$ is the ratio of the change between two lines drawn in the sample to the length before the sample is stretched.

Though the stress and strain are tensors, we turn all quantities into scalars by picking an experimental geometry in which all quantities change along the same direction.  As we see in Figure \ref{fig:Clamped-Unclamped}, the sample changes length along $x$, which is also the direction of the force.  As a result, all equations are kept in scalar form.

\begin{figure}[htbp]
   \centering
   \includegraphics{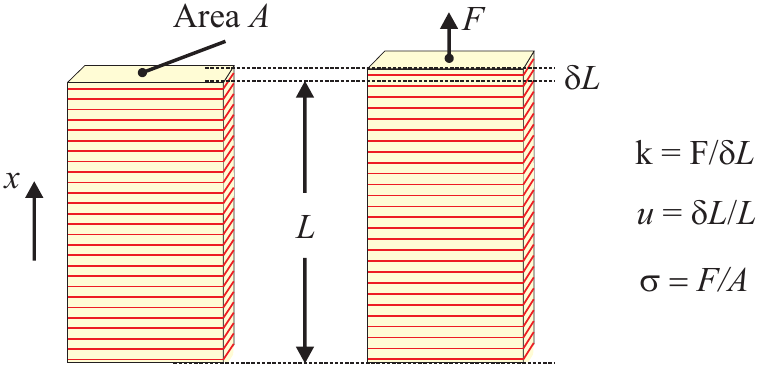}
   \caption{A force applied to the end of a sample causes it to expand like a spring.}
   \label{fig:Spring}
\end{figure}

Treating the material as a spring, an applied force $F$ yields an observed length change $\delta L$, so the spring constant is given simply by $k = F/x$.  Applying Equations \ref{eq:stress} and \ref{eq:strain} to the spring constant, we find
\begin{align}\label{eq:spring}
k = \frac {F} {\delta L} = \frac {\sigma A} {u L}.
\end{align}
\noindent Defining the Young's modulus $E$ by
\begin{align}\label{eq:young}
E = \frac {k L} {A},
\end{align}
\noindent Equation \ref{eq:spring} becomes
\begin{align}\label{eq:strss-strain}
\sigma = E u .
\end{align}
\noindent The spring constant depends on a material's geometry; $E$ is an intrinsic material property.

Like a bell, which continues to produce sound well after the ringer has struck, light can induce mechanical changes in a material that continue to act after the light is switched off.  This type of behavior is best represented by a response function, which relates the stress and strain at one time to the intensity at other times according to
\begin{align}\label{eq:StressResponse}
\sigma(t) = \int_{-\infty}^{\infty} dt^\prime \, \kappa_\sigma^{(1)} (t - t^\prime) I(t^\prime),
\end{align}
and
\begin{align}\label{eq:StrainResponse}
u(t) = \int_{-\infty}^{\infty} dt^\prime \, \kappa_u^{(1)} (t - t^\prime) I(t^\prime),
\end{align}
where $\kappa_\sigma^{(1)} (t - t^\prime)$ is the stress response function and  $\kappa_u^{(1)} (t - t^\prime)$ the strain response function.  Causality is built into the response function so that a stress or strain response cannot precede the light pulse.  Insofar as Equation \ref{eq:strss-strain} holds, the use of Equations \ref{eq:StressResponse} and \ref{eq:StrainResponse} yields
\begin{align}\label{eq:strss-strain-Reponse}
\kappa_\sigma^{(1)} (t - t^\prime) = E \kappa_u^{(1)} (t - t^\prime) ,
\end{align}
so we restrict the rest of the discussions to the stress response with the understanding that Equation \ref{eq:strss-strain-Reponse} can be used to determine the strain response function.

The response function can be determined most simply by applying an impulse stress of the form $I(t) = I_\delta \delta(t)$, which, from Equation \ref{eq:StressResponse}, yields
\begin{align}\label{eq:Delta-Reponse}
\kappa_\sigma^{(1)} (t) = \sigma(t)/I_\delta .
\end{align}
Thus, a measure of the stress as a function of time in response to a delta function is the most direct measure of the response function.

The response function can be highly messy, exhibiting different behavior on different time scales.  On the time scales relevant to the experiments presented here, polymeric materials are typically over-damped, so that the stress decays without oscillating.  As such, to a good approximation, the response function is given by the exponential
\begin{align}\label{eq:Damped-Reponse}
\kappa_\sigma^{(1)} (t) = \kappa_0^{(1)} \exp(-\alpha^{(1)} t) \theta(t),
\end{align}
where $\kappa_0^{(1)}$ is the strength of the photomechanical response, $\alpha^{(1)}$ is response rate, and $\theta(t)$ a step function centered at $t=0$ to enforce causality.  In the under-damped case, the response function oscillates.  With many processes acting on different time scales, the full response function is a sum of response functions of varying decay time constants and oscillation frequencies.

Experimentally, it is more convenient to apply a step function intensity rather than a delta function.  For the intensity function
\begin{align}\label{eq:StepIntenisty}
I = I_0 \theta(t),
\end{align}
and assuming that only one time constant is relevant, Equations \ref{eq:Damped-Reponse} and \ref{eq:StepIntenisty} substituted into Equation \ref{eq:StressResponse} yields
\begin{align}\label{eq:expt-stress-response-function}
\sigma (t) = \frac {\kappa_0^{(1)} I_0} {\alpha^{(1)}} \left( 1 - \exp(-\alpha^{(1)} t ) \right).
\end{align}
The measured time dependence of the stress is fit to Equation \ref{eq:expt-stress-response-function} to determine $\kappa_0^{(1)}$ and $\alpha^{(1)}$, which fully characterizes the response function through Equation \ref{eq:Damped-Reponse}.

Above we have derived the linear response function.  Generally, the stress response is a nonlinear function of the intensity and is given by
\begin{align}\label{eq:expt-stress-response-nonlinear}
\sigma (t) = \sum_{n=1}^\infty \frac {\kappa_0^{(n)} I_0^n} {\alpha^{(n)}} \left( 1 - \exp(-\alpha^{(n)} t ) \right),
\end{align}
where $\kappa_0^{(n)} $ is the $n^\text{th}$-order response amplitude.  If other mechanisms are present, labelled by index $m$, Equation \ref{eq:expt-stress-response-nonlinear} can be further generalized, yielding
\begin{align}\label{eq:expt-stress-response-nonlinear-mechs}
\sigma (t) = \sum_m \sigma^{(m)} (t).
\end{align}
As such, the student can tackle more general cases with a straightforward extension of the theory presented here.  It is simple to show that an exponential time dependence of the stress results also when the light is turned off.

We end this section by discussing the meaning of $\kappa_0^{(1)}$. Equation \ref{eq:Damped-Reponse} implies that
\begin{align}\label{eq:stress-response-t=0}
\kappa_0^{(1)} = \lim_{t \rightarrow 0} \kappa_\sigma^{(1)} (t)
\end{align}
is the instantaneous material response.  Taking the Fourier transform of Equation \ref{eq:Damped-Reponse} gives
\begin{align}\label{eq:Damped-Reponse-transform}
\tilde{\kappa}_\sigma^{(1)} (\omega) = \kappa_0^{(1)} \int_0^\infty d\omega \exp(-\alpha^{(1)} t) e^{i \omega t} = \frac {\kappa_0} {\alpha - i \omega}.
\end{align}
Then,
\begin{align}\label{eq:Damped-Reponse-zero-freq}
\lim_{\omega \rightarrow 0} \tilde{\kappa}_\sigma^{(1)} (\omega) = \frac {\kappa_0} {\alpha} \equiv \tilde{\kappa}.
\end{align}
Therefore
\begin{align}\label{eq:kappa-kappa-tilde}
\tilde{\kappa} = \frac {\kappa_0} {\alpha}
\end{align}
is the static part of the response function in frequency space.  In SI, $\kappa$ has units of $m^{-1}$ and $\tilde{\kappa}$ has units $s \cdot m^{-1}$. 

\section{Weighted Averages}\label{sec:weightedavg}

We suggest collecting ten on-off cycles for each intensity. After fitting each cycle, 10 fit constants are collected. In general, one may perform an average on these fit constants, but a weighted average is preferred. A weighted average uses each datum's error (specifically, error$^{-1}$) as a weighting factor. As such, a datum with a large error contributes less to the resultant average than a datum with a smaller error.

Fit constants may be averaged using a weighted average,
\begin{eqnarray}
\bar{\sigma} \left(I\right)&=& \frac{\sum_{i=0}^{n}\sigma^{\mathrm{on}}_{eq,i}\left(I\right)/\Delta \sigma^{\mathrm{on}}_{eq,i}\left(I\right)}{\sum_{i=0}^n 1/\Delta \sigma^{\mathrm{on}}_{eq,i}\left(I\right)}
\end{eqnarray}
with associated statistical uncertainty
\begin{eqnarray}
\Delta \bar{\sigma} & =& \sqrt{\frac{1}{\sum_{i=0}^n 1/ \left(\Delta \sigma^{\mathrm{on}}_{eq,i}\left(I\right) \right)^2}}.
\end{eqnarray}
\end{document}